\documentclass[footinbib,aip,apl,preprint,longbibliography]{revtex4-1}

\usepackage[absolute]{textpos}
\usepackage{bbm}
\usepackage{verbatim}
\usepackage{graphicx}
\usepackage{amsmath}
\usepackage{bm}

\begin{document}
\title{Controllable two-photon interference with versatile quantum frequency processor}

\author{Hsuan-Hao Lu}
\thanks{These authors contributed equally to this work.}
\affiliation{School of Electrical and Computer Engineering and Purdue Quantum Center, Purdue University, West Lafayette, Indiana 47907, USA}
\author{Joseph M. Lukens}
\thanks{These authors contributed equally to this work.}
\affiliation{Quantum Information Science Group, Computational Sciences and Engineering Division, Oak Ridge National Laboratory, Oak Ridge, Tennessee 37831, USA}
\author{Nicholas A. Peters}
\affiliation{Quantum Information Science Group, Computational Sciences and Engineering Division, Oak Ridge National Laboratory, Oak Ridge, Tennessee 37831, USA}
\affiliation{Bredesen Center for Interdisciplinary Research and Graduate Education, The University of Tennessee, Knoxville, Tennessee 37996, USA}
\author{Brian P. Williams}
\affiliation{Quantum Information Science Group, Computational Sciences and Engineering Division, Oak Ridge National Laboratory, Oak Ridge, Tennessee 37831, USA}
\author{Andrew M. Weiner}
\affiliation{School of Electrical and Computer Engineering and Purdue Quantum Center, Purdue University, West Lafayette, Indiana 47907, USA}
\author{Pavel Lougovski}
\affiliation{Quantum Information Science Group, Computational Sciences and Engineering Division, Oak Ridge National Laboratory, Oak Ridge, Tennessee 37831, USA}
\date{\today}

\begin{textblock}{13.45}(1.35,12)
\noindent \fontsize{7}{7}\selectfont This manuscript has been authored by UT-Battelle, LLC under Contract No. DE-AC05-00OR22725 with the U.S. Department of Energy. The United States Government retains and the publisher, by accepting the article for publication, acknowledges that the United States Government retains a non-exclusive, paid-up, irrevocable, world-wide license to publish or reproduce the published form of this manuscript, or allow others to do so, for United States Government purposes. The Department of Energy will provide public access to these results of federally sponsored research in accordance with the DOE Public Access Plan. (http://energy.gov/downloads/doe-public-access-plan).
\end{textblock}

\maketitle

\textbf{Quantum information is the next frontier in information science, promising unconditionally secure communications, enhanced channel capacities, and computing capabilities far beyond their classical counterparts~\cite{Gisin2002, Ladd2010}. And as quantum information processing devices continue to transition from the lab to the field, the demand for the foundational infrastructure connecting them with each other and their users---the quantum internet~\cite{Kimble2008, Pirandola2016}---will only increase. Due to the remarkable success of frequency multiplexing and control in the \emph{classical} internet, quantum information encoding in optical frequency offers an intriguing synergy with state-of-the-art fiber-optic networks. Yet coherent quantum frequency operations prove extremely challenging, due to the difficulties in mixing frequencies efficiently, arbitrarily, in parallel, and with low noise. Here we implement an original approach based on a reconfigurable quantum frequency processor, designed to perform arbitrary manipulations of spectrally encoded qubits. This processor's unique tunability allows us to demonstrate frequency-bin Hong-Ou-Mandel interference with record-high 94\% visibility. Furthermore, by incorporating such tunability with our method's natural parallelizability, we synthesize independent quantum frequency gates in the same device, realizing the first high-fidelity flip of spectral correlations on two entangled photons. Compared to quantum frequency mixing approaches based on nonlinear optics~\cite{Kobayashi2016, Clemmen2016}, our linear method removes the need for additional pump fields and significantly reduces background noise. Our results demonstrate multiple functionalities in parallel in a single platform, representing a huge step forward for the frequency-multiplexed quantum internet.}

In the classical domain, the ultrabroad bandwidth supported by optical fiber has proven crucial in solidifying fiber optics in the digital communications revolution. Wavelength-division multiplexing (WDM)---either in its standard embodiment with independent frequency channels, or more complex versions with channels comprising interleaved bands~\cite{Hillerkuss2011}---forms an essential component, and will continue to do so even as novel data formats and multiplexing techniques are incorporated~\cite{Agrell2016}. Such success has naturally brought WDM approaches to the forefront for the \emph{quantum} internet as well. In particular, potentially large amounts of information can be stored in single photons encoded in spectro-temporal modes~\cite{Lukens2014, Brecht2015, Zhong2015, Islam2017}, and frequency multiplexing is essential for scaling up quantum memories~\cite{Sinclair2014}.

To that end, considerable progress has been made in generating multiphoton entanglement across a comb of narrowband frequency modes, or bins, including optical parametric oscillators below threshold~\cite{Lu2003}, straightforward filtering of broadband parametric downconversion~\cite{Xie2015}, and, recently, on-chip production of quantum frequency combs using microring resonators~\cite{Reimer2016, Jaramillo2017, Kues2017, Imany2018}. Likewise, an explosion of research in quantum frequency conversion has showcased coherent translation of single-photon states across both wide~\cite{Huang1992, Tanzilli2005} and narrow~\cite{Wright2017} bandwidths. But the step from viewing frequency as a channel---wherein quantum information is carried by some other parameter, such as time or polarization---to encoding the information itself in frequency is significant, requiring markedly more complex operations: i.e., universal gate sets in frequency modes. As important milestones in that direction, frequency beamsplitters~\cite{Kobayashi2016, Clemmen2016, Joshi2017b} and quantum pulse gates~\cite{Brecht2015}
% I've really truncated QPG references, but I don't want to get pulled down that path very far. Is this ok? %
 based on optical nonlinearities have shown coherent interference and mode selection of frequency-encoded photons. Yet the need for powerful optical control fields makes nonlinear approaches challenging, given the potential for extra noise photons from Raman scattering and imperfect isolation.

\begin{figure*}
\centering
\includegraphics[width=4.5in,trim={0in 0in 0.0in 0in},clip]{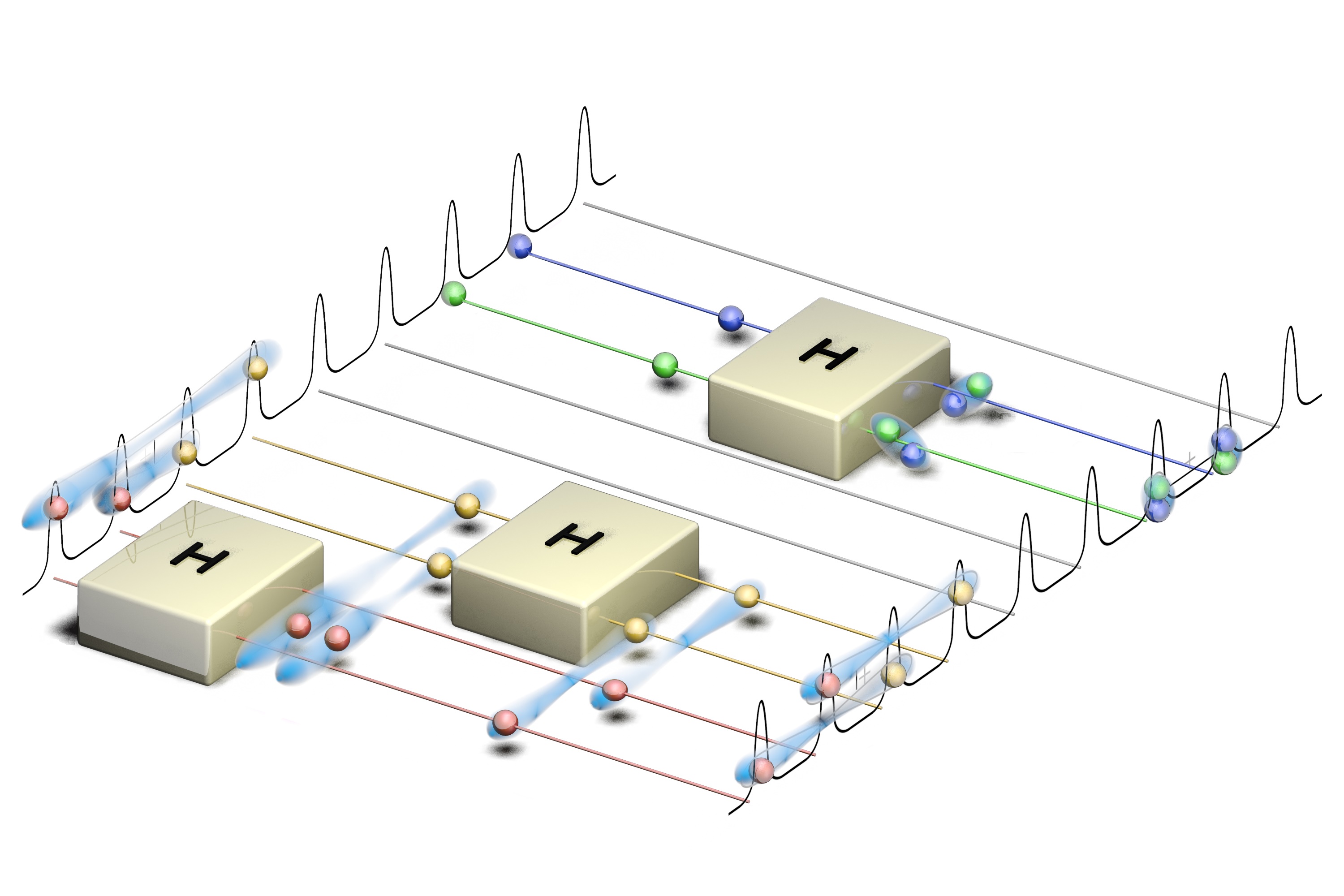}
\caption{\textbf{High-level vision of quantum frequency processor.} Single photons (spheres) populating a comb of frequency bins propagate through a parallelized network of quantum gates (boxes) performing the desired set of operations. Spheres of a specific color trace the probability amplitudes of a single input photon, so that an ideal measurement will register precisely one click for each color. Frequency superpositions are represented by spheres straddling multiple lines, while entangled states are sums of photon products (visualized by clouds). The specific operations are those we realize experimentally: Hong-Ou-Mandel interference (top) and two-qubit rotation (bottom).}
\label{fig1}
\end{figure*}

Recently, we proposed a general framework~\cite{Lukens2017} for spectrally encoded photonic state control, based on electro-optic phase modulators (EOMs) and Fourier-transform pulse shapers. Enabling universal quantum information processing in a scalable fashion, our approach is also optically linear, obviating the need for additional pump fields. Figure~\ref{fig1} sketches an example of such a quantum frequency processor,
% In the figures we sometimes use the abbreviation QFP, but I like avoiding it in the text as much as possible for flow...hopefully not too confusing. %
 with the particular operations chosen to match the ensuing experiments. In general, an input quantum state consisting of a superposition of photons spread over discrete frequency bins is manipulated by the designed network of EOMs and pulse shapers, which applies various unitary operations to combinations of frequency bins. After each step, some of the frequency bins can be detected, with the newly available bandwidth re-provisioned with freshly encoded photons. Note that, although we draw each frequency bin as a separate ``rail'' for conceptual purposes, the physical encoding occurs within a single fiber-optic spatial mode, thereby enabling natural phase stability and providing compatibility with current fiber networks. This paradigm has allowed us to experimentally demonstrate frequency beamsplitters and tritters with ultrahigh operation fidelity and parallelizability across a 40-nm optical bandwidth~\cite{Lu2018}. While then validated with weak coherent states, a quantum frequency network will require multiphoton and nonclassical interference phenomena as well. In this work, we experimentally show that not only does our operation indeed realize unrivaled interference of fully quantum frequency states, but it also enables independent simultaneous operations \emph{in the same device}, via its unique tunability and parallelizability.

Two nonclassical phenomena of particular significance in quantum photonics are Hong-Ou-Mandel (HOM) interference~\cite{Hong1987} and the Einstein-Podolsky-Rosen (EPR) paradox~\cite{Einstein1935}. In the conventional HOM interferometer, two photons mixed on a 50/50 spatial beamsplitter bunch, never exiting in different output ports; a general feature of bosons, HOM interference forms the basis of essentially all multiqubit gates in linear optics~\cite{Kok2007}. The EPR paradox notes the strange behavior in quantum mechanics that two spatially distant particles can share strong correlations in noncommuting observables (such as position and momentum). This highlights a fundamental inconsistency between the completeness of quantum mechanics and local realism, ultimately underpinning Bell tests of nonlocality~\cite{Bell1964} and security in quantum key distribution~\cite{Gisin2002}. Both phenomena are properly foundational in that they shed light on fundamental interpretations of quantum mechanics and enable practical quantum information applications.

\begin{figure}
\includegraphics[width=4.5in]{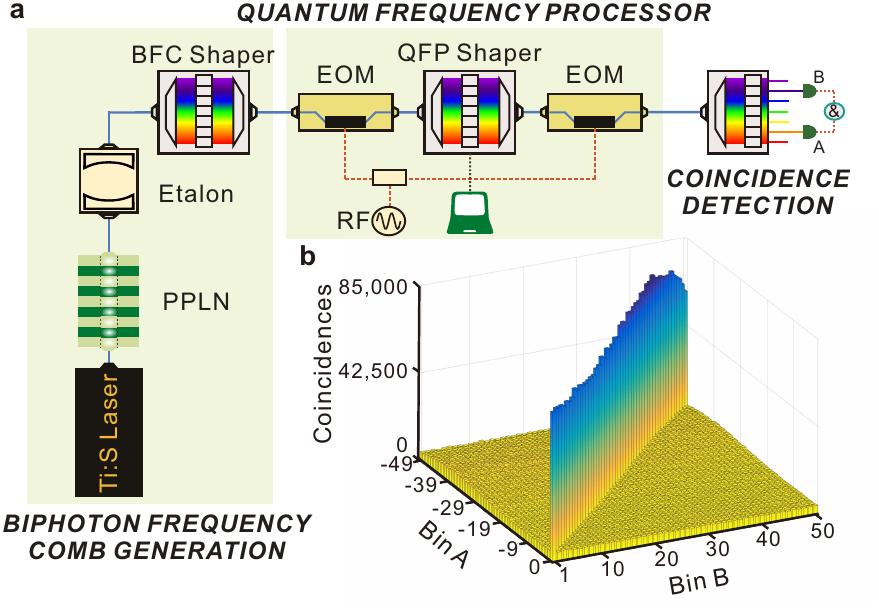}
\caption{\textbf{Quantum frequency processor setup.} \textbf{a}, Experimental configuration. Parametric downconversion followed by an etalon generates a biphoton frequency comb with 25-GHz spacing, which is spectrally filtered and sent through the quantum frequency processor, after which a wavelength-selective switch is used to measure photon coincidences between frequency bins. \textbf{b}, Joint frequency spectrum of source, measured with etalon output connected directly to coincidence detection setup. Frequency bins satisfying $n_A+n_B=1$ show strong correlations, whereas all other combinations are at the expected accidentals level. Coincidences are counted over 5 seconds.}
\label{fig2}
\end{figure}

Figure~\ref{fig2}a shows our setup for processing quantum information encoded in frequency. Our test source of entanglement is a biphoton frequency comb (BFC) generated by pumping a periodically poled lithium niobate (PPLN) waveguide with a continuous-wave Ti:sapphire laser, filtering the broadband emission with an etalon to produce 25-GHz-spaced frequency bins, and setting the relative amplitude and phase of each frequency bin with a pulse shaper. In Fig.~\ref{fig2}b, we plot the measured frequency correlations of this source, obtained by bypassing the quantum frequency processor (QFP), scanning the filters of the output wavelength-selective switch, and counting coincidences between two detectors. Each frequency-bin index $n$ corresponds to the filter centered at $\omega_n = \omega_0 + n\Delta\omega$, where $\omega_0/2\pi = 193.6000$ THz (ITU channel 36 at 1548.51 nm) and $\Delta\omega/2\pi = 25$ GHz. Over this $50\times 50$ mode grid, we observe high coincidence counts only for frequency-bin pairs satisfying $n_A+n_B=1$, as expected by energy conservation. The processor itself consists of a pulse shaper sandwiched between two EOMs. Each EOM is driven by a 25-GHz sinusoidal voltage, while the pulse shaper imparts a user-defined phase to each spectral bin; this combination was shown to enable a frequency Hadamard gate $H$ with 99.998\% fidelity and only 2.61\% photon leakage into neighboring modes~\cite{Lu2018}.

\begin{figure}
\centering
\includegraphics[width=6in,trim={1in 0in 1in 0in},clip]{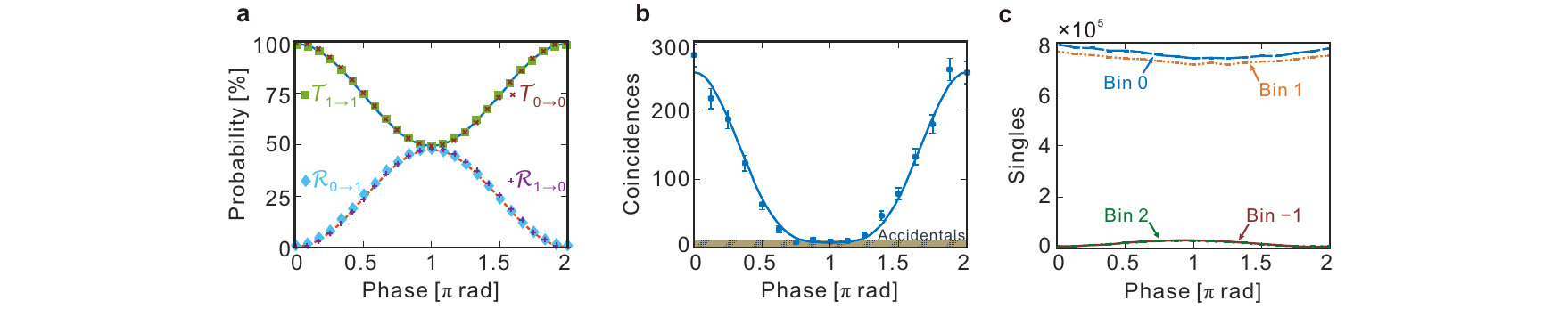}
\caption{\textbf{Frequency HOM interference.} \textbf{a}, Beamsplitter reflectivities $\mathcal{R}$ and transmissivities $\mathcal{T}$ for all paths between frequency bins 0 and 1, as pulse shaper phase shift is tuned. Markers denote the values measured with a coherent state probe, while curves give the theory. \textbf{b}, Measured output coincidence counts between bins 0 and 1, for a two-photon input (no accidentals subtraction). The HOM visibility is $0.94\pm 0.01$. \textbf{c}, Singles counts in bins 0, 1, and adjacent $-1$ and $2$ for the same input as \textbf{b}. Here detector dark counts are subtracted to compare output flux. For \textbf{b} and \textbf{c}, counts are recorded over 180 s, and error bars assume Poissonian statistics.}
\label{fig3}
\end{figure}

In the case of HOM interference, we seek to apply such a gate to a pair of photons located in adjacent frequency bins, which we obtain directly by filtering out all but bins 0 and 1 of the source (Fig.~\ref{fig2}b);  the $H$ gate applied to 0 and 1 should cause both photons to bunch in either bin 0 or 1, with no coincidences between the two bins. To measure the strength of quantum interference, one must scan some parameter which controls the distinguishability of the two-photon probability amplitudes leading to clicks on both output detectors; a visibility exceeding $50\%$ indicates nonclassicality~\cite{Hong1987}. In the case of frequency mixers, one can introduce a temporal delay between the two modes~\cite{Kobayashi2016} or scan the photon frequency spacing relative to that of the frequency beamsplitter~\cite{Joshi2017b}. In our case, we adjust the mixing probability of the operation itself, analogous to varying the reflectivity $\mathcal{R}$ of a spatial beampslitter. To do so, we note a valuable feature of the our spectral beamsplitter: tunability. For by simply changing the depth of the phase shift imparted by the pulse shaper between frequency bins 0 and 1, the spectral reflectivity $\mathcal{R}$ can be tuned smoothly from 0 to $\sim$0.5 and back to 0 (see Methods). Figure~\ref{fig3}a plots the theoretically predicted (curves) and experimentally measured (symbols) beamsplitter transmission and reflection coefficients between bins 0 and 1, when scanning the pulse shaper phase. A phase setting of $\pi$ results in an $H$ gate; $0$ and $2\pi$ phase shifts yield an identity operation. It is important to note that both EOMs remain fixed throughout the scan, so that the tunability is effected only by adjusting the phase applied by the pulse shaper.

Sending in the photon pair $|1_{\omega_0}\rangle_A |1_{\omega_1}\rangle_B$ (i.e., one photon in frequency-bin 0, assigned to party $A$, and one photon in frequency-bin 1, assigned to party $B$) and scanning the pulse shaper phase, we measure the coincidence counts between output bins 0 and 1 shown in Fig.~\ref{fig3}b. The solid curve is the theoretical prediction, scaled and vertically offset to match the data points via linear least squares; the visibility obtained from this fit is $0.94 \pm 0.01$, with the reduction from unity completely consistent with the accidentals level expected for our measured counts and timing resolution. This visibility far exceeds the previous values measured for frequency-domain HOM interference---namely, $0.71 \pm 0.04$~\cite{Kobayashi2016} and $0.68\pm 0.03$~\cite{Joshi2017b}---and is a consequence of both the reduced optical noise and fine controllability of the operation. We also record the singles counts for bins $0$ and $1$, as well as the adjacent sidebands ($-1$ and $2$). As shown in Fig.~\ref{fig3}c, the two central modes retain nearly constant flux across the full scan, showing that the dip in coincidence counts results from truly quantum HOM interference as opposed to photon loss (see Methods for detailed verification).
% [Indeed, refitting to the singles-normalized $g_{01}^{(2)}$ function between modes 0 and 1 still retrieves a visibility of $0.94 \pm 0.01$, confirming such intuition (see Methods).] %
 Moreover, the small reduction in singles counts around $\pi$---accompanied by the increase in singles counts for bins $-1$ and $2$---also qualitatively matches expectations, given the fact that the full $H$ gate scatters $2.61\%$ of the input photons out of the computational space into adjacent sidebands. We note that even this scattering could be  removed by driving the EOMs with more complicated waveforms~\cite{Lukens2017}. 

\begin{figure}
\includegraphics[width=4.5in]{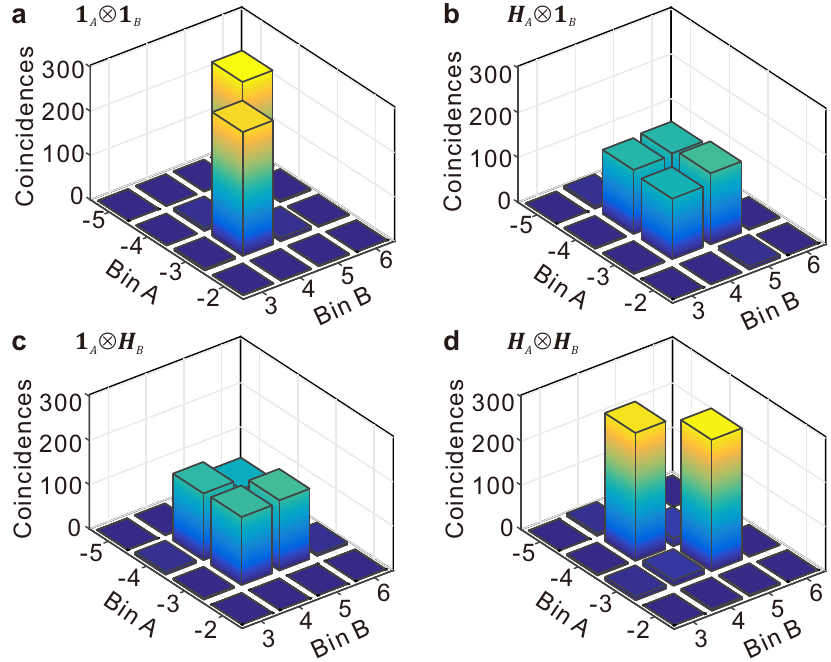}
\caption{ \textbf{Two-photon state manipulation.} Coincidences between output frequency bins after application of the following gates: \textbf{a}, identity on both photons; \textbf{b}, Hadamard on photon $A$; \textbf{c}, Hadamard on photon $B$; \textbf{d}, Hadamard on both photons. Coincidences are collected over 120 s.}
\label{fig4}
\end{figure}

The quantum frequency processor's tunability, invoked in the above realization of HOM interference, relies only on modifying the spectral phase, which suggests the ability to perform independent operations by setting different phase shifts on appropriate subbands in the pulse shaper's bandwidth. Accordingly, this form of parallelizability is even stronger than previously shown, where the \emph{same} operation was replicated across the bandwidth~\cite{Lu2018}. To demonstrate this, we set the BFC shaper to pass modes $\{-4, -3, 4, 5\}$ (cf. Fig.~\ref{fig2}b), preparing the input entangled state $|\Psi\rangle \propto |1_{\omega_{-4}}\rangle_A |1_{\omega_5}\rangle_B + |1_{\omega_{-3}}\rangle_A |1_{\omega_4}\rangle_B$. On each pair of frequency bins---$\{-4,-3\}$ and $\{4,5\}$---we set the spectral phase to apply either the identity $\mathbbm{1}$ or Hadamard $H$ gates, and then measure coincidence counts between the frequency bins at the output. Figure~\ref{fig4} furnishes the results for all four combinations of $\mathbbm{1}$ and $H$; when the two gates match, near-perfect spectral correlations result (a and d), whereas mismatched cases produce uniform population of the two-qubit space (b and c). By measuring correlations in adjacent bins as well, we confirm the self-contained nature of our operation; even in the worst case (Fig.~\ref{fig4}d), less than $6\%$ of the total coincidences lie outside of the $2\times 2$ subspace, whereas similar state manipulation with only one EOM suffers from high probability of qubit scattering~\cite{Kues2017, Imany2018}. Importantly, the transition from $\mathbbm{1}_A \otimes \mathbbm{1}_B$ to $H_A \otimes H_B$ actually flips the correlations entirely, eliminating the negative frequency dependence resulting from pump energy conservation in favor of a positive dependence.  Our demonstration is the first sign flip of entangled-photon frequency correlations \emph{post-generation} and the first electro-optic-based two-qubit rotation that is closed in the $2\times 2$ computational space---essential for the ideal two-level logic underpinning qubit-based forms of quantum information processing.

Using the strong correlations in two mutually unbiased measurements ($\mathbbm{1}$ and $H$), we can quantify directly the EPR-like nature of our entangled state. We define the conditional entropies $\mathcal{H}(\mathbbm{1}_A|\mathbbm{1}_B)$ and $\mathcal{H}(H_A|H_B)$ as the uncertainty of the measured frequency mode of $A$ \{$-4,-3$\} given knowledge of $B$'s result \{$4,5$\}, for the two cases of matched transformations (Fig.~\ref{fig4}a and d). Retrieving the probabilities from the raw counts via Bayesian mean error estimation (with no accidental subtraction), we recover $\mathcal{H}(\mathbbm{1}_A|\mathbbm{1}_B)= 0.19 \pm 0.03$ and $\mathcal{H}(H_A|H_B) = 0.29 \pm 0.04$, violating the Maassen-Uffink bound $q_{MU}$ for separable states [$\mathcal{H}(\mathbbm{1}_A|\mathbbm{1}_B) +  \mathcal{H}(H_A|H_B) \geq q_{MU}$]~\cite{Coles2017} by nearly 10 standard deviations (see Methods). Moreover, despite the fact we have implemented a tomographically incomplete set of measurements in Fig.~\ref{fig4}, Bayesian methods allow us to estimate the full density matrix; any lack of information is reflected directly in the retrieved uncertainty. Doing so, we obtain the density matrix $\hat{\rho}$ with fidelity $\mathcal{F}=\langle\Psi|\hat{\rho}|\Psi\rangle = 0.92\pm0.01$, where $|\Psi\rangle$ is the maximally entangled ideal state (see Methods and Extended Data Fig.~1). Such findings demonstrate the utility of our quantum frequency processor for manipulating joint quantum systems coherently and independently, preserving a state's built-in entanglement in the process---an entirely new functionality in frequency-bin qubit control.

\textbf{Acknowledgements.} We thank A.~Sproles for the graphics in Fig.~\ref{fig1}, D.~E. Leaird for technical assistance, and B.~Qi for valuable discussions. This work was performed in part at Oak Ridge National Laboratory, operated by UT-Battelle for the U.S. Department of Energy under Contract No. DEAC05-00OR22725. Funding was provided by ORNL's Laboratory Directed Research and Development Program.

\textbf{Author Contributions.} H-H.L. led the experiments and data synthesis, and contributed to the theoretical analysis. J.M.L. initiated the concepts, assisted in the laboratory, and led writing of the paper. N.A.P. assisted with experiments, data interpretation, and conceptual design. B.P.W. set up the time-tagging electronics and performed the Bayesian estimation. A.M.W. supervised the work at Purdue and assisted with system design. P.L. initiated the concepts, managed the theoretical analysis, and supervised the project. All authors reviewed the results and contributed to the manuscript.

\textbf{Author Information.} The authors declare no competing financial interests. Correspondence and requests for materials should be addressed to J.M.L.~(lukensjm@ornl.gov) or P.L.~(lougovskip@ornl.gov).

%%%%%%%%%%%%%%%%%%%%%%%%%%%%%%%%%%%%%%%%%%%%%%%%%%%%%%%%%%%%%%%%%%%%%%%%%%%%%%%%%%%%%%%%%%%%
%\newpage
\section*{METHODS}
\textbf{Experimental details.}
We couple a continuous-wave Ti:sapphire laser (M Squared) into a fiber-pigtailed periodically poled lithium niobite (PPLN; SRICO) waveguide, temperature controlled at $\sim$85$^{\circ}$C for spontaneous parametric down-conversion under type-0 phase matching. Spectrally entangled photon pairs spanning $>$2.5 THz are subsequently filtered by a Fabry-Perot etalon (Optoplex) with 25-GHz mode spacing (matched to the ITU grid) to produce a biphoton frequency comb (BFC), with each comb line possessing a full-width at half-maximum linewidth of 1.8 GHz. The center frequency of the pump laser is carefully locked to align the generated signal-idler pairs with etalon peaks, i.e., to maximize coincidences between the spectrally filtered modes. We utilize a pulse shaper (BFC shaper; Finisar) to perform amplitude and phase filtering to prepare particular input states for quantum frequency processing. Our quantum frequency processor (QFP) consists of two 40-Gb/s EOMs (EOSpace) with a pulse shaper~\cite{Weiner2011} (QFP shaper; Finisar) sandwiched between them, with a total insertion loss of 12.5 dB~\cite{Lu2018}. We note that synchronizing the biphoton emission time to the EOM phase is not necessary when the two-photon coherence time is much larger than the modulation period. This condition is guaranteed, e.g., by a pump laser with narrow linewidth, as in the current experiments. To implement a Hadamard operation, i.e., frequency-bin beamsplitter, we drive the two EOMs with 25-GHz $\pi$-phase-shifted sinewaves, and apply a step function with $\pi$-phase jump between the two computational modes on the shaper. The specific phase patterns are obtained in advance from optimization program in refs.~\onlinecite{Lukens2017,Lu2018}, which achieve fidelity $\mathcal{F} =0.9999$ and success probability $\mathcal{P} =0.9760$ numerically for the Hadamard operation. The output photons are frequency-demultiplexed by an amplitude-only wavelength selective switch (WSS; Finisar) having 12.5-GHz channel specificity.  Each time we route two different spectral modes (each takes up two pixels on the WSS) to two superconducting nanowire single-photon detectors (SNSPD; Quantum Opus) to record single counts as well as the coincidences within 1.5-ns bins. 

\textbf{Quantum operations.} The specific configuration for our Hadamard gate (cf. supplement of ref.~\onlinecite{Lu2018}) relies on the temporal phase modulation $\varphi(t) = \pm \Theta \sin\Delta\omega t$ ($\Theta = 0.8169$ rad) on the first and second EOMs, respectively. And for a gate operating on bins 0 and 1, the discrete pulse shaper phases can be written as
\begin{equation}
\label{E0}
\nonumber
\phi_n =
\begin{cases}
\phi_0 & ; n \leq 0 \\
\phi_0 + \alpha & ; n \geq 1.
\end{cases}
\end{equation}
Here $\phi_0$ is an offset with no physical significance, while $\alpha=\pi$ for the ideal Hadamard. Yet $\alpha$ can be tuned as well; doing so actually permits tunable reflectivity. 
\begin{comment}
Specifically, numerical simulations show that we can approximate the $2\times 2$ transformation matrix on bins 0 and 1 as
\begin{equation}
\label{E2}
\nonumber
V = 
\begin{bmatrix}
|J_0(\Theta)|^2+\sum_{i=1}^\infty |J_i(\Theta)|^2 (1+e^{j\alpha})  & \sum_{i=1}^\infty |J_i(\Theta) J_{i-1}(\Theta)| (1-e^{j\alpha}) \\
\sum_{i=1}^\infty |J_i(\Theta) J_{i-1}(\Theta)| (1-e^{j\alpha})  & |J_0(\Theta)|^2 (e^{j\alpha})+\sum_{i=1}^\infty |J_i(\Theta)|^2 (1+e^{j\alpha})
\end{bmatrix}
\end{equation}
\end{comment}
Specifically, if we write out the $2\times 2$ transformation matrix on modes 0 and 1 as a function of this phase,
\begin{equation}
\label{E3}
\nonumber
V =
\begin{bmatrix}
V_{00}(\alpha) & V_{01}(\alpha) \\
V_{10}(\alpha) & V_{11}(\alpha)
\end{bmatrix},
\end{equation}
we can define the variable reflectivities (i.e., mode-hopping probabilities) and transmissivities (probabilities of preserving frequency) as
\begin{equation}
\nonumber
%\begin{aligned}
%\mathcal{R}_{0\rightarrow 1} = |V_{10}(\alpha)|^2 = 4 \left[ \sum_{k=1}^\infty J_k(\Theta) J_{k-1}(\Theta)  \right]^2 \sin^2\frac{\alpha}{2}\\
%\mathcal{R}_{1\rightarrow 0} = |V_{01}(\alpha)|^2 = 4\left[ \sum_{k=1}^\infty J_k(\Theta) J_{k-1}(\Theta)  \right]^2 \sin^2\frac{\alpha}{2} \\
%\mathcal{T}_{0\rightarrow 0} = |V_{00}(\alpha)|^2 = J_0^4(\Theta) \sin^2\frac{\alpha}{2} + \cos^2\frac{\alpha}{2} \\
%\mathcal{T}_{1\rightarrow 1} = |V_{11}(\alpha)|^2 = J_0^4(\Theta) \sin^2\frac{\alpha}{2} + \cos^2\frac{\alpha}{2},
%\end{aligned}
\begin{aligned}
\mathcal{R}_{0\rightarrow 1} = |V_{10}(\alpha)|^2 = \left| (1-e^{i\alpha}) \sum_{k=1}^\infty J_k(\Theta) J_{k-1}(\Theta)  \right|^2 \\
\mathcal{R}_{1\rightarrow 0} = |V_{01}(\alpha)|^2 = \left| (1-e^{i\alpha}) \sum_{k=1}^\infty J_k(\Theta) J_{k-1}(\Theta)  \right|^2 \\
\mathcal{T}_{0\rightarrow 0} = |V_{00}(\alpha)|^2 = \left| J_0^2(\Theta) + (1+e^{i\alpha})\frac{1-J_0^2(\Theta)}{2} \right|^2 \\
\mathcal{T}_{1\rightarrow 1} = |V_{11}(\alpha)|^2 = \left| e^{i\alpha} J_0^2(\Theta) + (1+e^{i\alpha})\frac{1-J_0^2(\Theta)}{2} \right|^2,
\end{aligned}
\end{equation}
where $J_k(\Theta)$ is the Bessel function of the first kind. We note that, when $\alpha=\pi$, the elements $\{V_{00},V_{01},V_{10}\}$ are all real and positive, while $V_{11}$ is real and negative---in accord with the ideal Hadamard and leading to destructive HOM interference between the reflect/reflect and transmit/transmit two-photon probability amplitudes. Additionally, these expressions satisfy $\mathcal{R}_{0\rightarrow 1}=\mathcal{R}_{1\rightarrow 0}\equiv \mathcal{R}$ and $\mathcal{T}_{0\rightarrow 0} = \mathcal{T}_{1\rightarrow 1} \equiv \mathcal{T}$. As $\alpha$ is tuned over $0\rightarrow\pi\rightarrow2\pi$, $\mathcal{R}$ follows from 0 to a peak of 0.4781 and back to 0, while $\mathcal{T}$ starts at 1, drops to 0.4979, and returns to 1. The sum $\mathcal{R+T}$ defines the gate success probability, which drops slightly at $\alpha=\pi$ due to the use of single-frequency electro-optic modulation. These particular values are confirmed experimentally in Fig.~\ref{fig2}a with coherent state measurements~\cite{Lu2018}.

\textbf{Hong-Ou-Mandel interference.}
The generated biphoton frequency comb can be described as a state of the form
\begin{equation}
\label{e1}
|\Psi\rangle = \sum_{n=1}^N c_n |1_{\omega_{1-n}}\rangle_A  |1_{\omega_{n}}\rangle_B,  \end{equation}
or in terms of bosonic mode operators,
\begin{equation}
\label{e1-1}
\nonumber
|\Psi\rangle = \sum_{n=1}^N c_n \hat{a}_{1-n}^\dagger \hat{a}_{n}^\dagger |\mathrm{vac}\rangle_A |\mathrm{vac}\rangle_B,
\end{equation}
where $\hat{a}_{n}$ ($\hat{a}_{n}^\dagger$) annihilates (creates) one photon in the frequency bin centered at $\omega_n$. The $A$ and $B$ nomenclature defines the modes held by each of two parties: $A$ consists all $\omega_n$ such that $n\leq 0$, $B$ everything with $n\geq1$. We favor this notation over the more traditional ``signal'' and ``idler'' classification because (i) our frequency operations can move photons between $A$ and $B$ mode sets---and indeed \emph{does} in the case of HOM---and (ii) there are no other distinguishing degrees of freedom to label the photons.

Our quantum frequency processor transforms these bins into outputs $\hat{b}_m$ (at frequencies $\omega_m$) via
\begin{equation}
\label{e2}
\nonumber
\hat{b}_m = \sum_{n=-\infty}^\infty V_{mn} \hat{a}_n.
\end{equation}
The matrix $V$ describes the entire operation over all modes. Then at the output we measure the spectrally resolved coincidences between bins $n_A$ and $n_B$, i.e.,
\begin{equation}
\label{e3}
\nonumber
C_{n_A n_B} = \langle\Psi |\hat{b}_{n_A}^\dagger \hat{b}_{n_B}^\dagger \hat{b}_{n_B} \hat{b}_{n_A}| \Psi\rangle,
\end{equation}
as well as the singles
\begin{equation}
\label{e4}
\nonumber
S_n = \langle\Psi |\hat{b}_{n_A}^\dagger \hat{b}_{n_A}| \Psi\rangle
\end{equation}

In the case of HOM interference, we filter out all photon pairs except $c_1$ [Eq.~(\ref{e1})], so the input state is $|\Psi\rangle = |1_{\omega_0}\rangle_A |1_{\omega_1}\rangle_B$, which gives $C_{01} = |V_{00}V_{11}+V_{01}V_{10} |^2$ and $S_n= |V_{n0}|^2+|V_{n1}|^2$. In light of the previous discussion on beamsplitter tunability, we thus predict:
\begin{equation}
\label{e5}
\nonumber
\begin{aligned}
C_{01}=|\mathcal{R}(\alpha)-\mathcal{T}(\alpha)|^2 \\
S_0 = S_1 = \mathcal{R}(\alpha) + \mathcal{T}(\alpha) \\
S_{-1} = S_2 \approx 1 - \mathcal{R}(\alpha) - \mathcal{T}(\alpha),
\end{aligned}
\end{equation}
where the nonunity success probability [$\mathcal{R}(\pi)+\mathcal{T}(\pi)=0.976$] results in some photons scattering into bins $-1$ and $2$. (Scattering beyond these modes is not observable in experiment, consistent with the theoretical prediction of only $\sim$10$^{-4}$ probability to leave the center four bins.) Invoking the theoretically predicted values for $\mathcal{R}$ and $\mathcal{T}$, we use weighted least-squares to fit the function $f(\alpha)=K_0+K_1 C_{01}(\alpha)$ to the data in Fig.~\ref{fig3}b and extract the visibility
\begin{equation}
\label{e51}
\nonumber
\mathcal{V}=\frac{K_1 \left[C_{01}(0)-C_{01}(\pi) \right]}{2K_0+K_1 \left[C_{01}(0)+C_{01}(\pi) \right]} = 0.94\pm 0.01.
\end{equation}
Now, because the singles $S_0$ and $S_1$ drop slightly at $\alpha=\pi$ (cf. Fig.~\ref{fig3}b)---which is not the case in a traditional HOM experiment---we also look at the visibility of the normalized cross-correlation function, $g_{01}^{(2)} = \frac{C_{01}}{S_0 S_1}.$ For in the most pathological case, a reduction in the unnormalized coincidences $C_{01}$ could in principle be due to dropping singles $S_0$ or $S_1$, which would not be surprising from a classical view: if one detector rarely clicks, of course its coincidences with another detector will drop as well. On the other hand, the normalized $g_{01}^{(2)}$ does not suffer from this issue, by accounting for singles counts directly. Accordingly, we repeat the least-squares fit using the theoretically predicted $g_{01}^{(2)}(\alpha)$, along with the measured coincidences (Fig.~\ref{fig3}b) divided by the product of mode 0 and 1 single counts (Fig.~\ref{fig3}c). In this more conservative case, we still retrieve $\mathcal{V}=0.94 \pm 0.01$, fully confirming the nonclassicality of our HOM interference.

\textbf{Quantum state manipulation.} For the state rotation experiments, we filter out all modes except four, leaving the entangled qubits [$n=4,5$ in Eq.~(\ref{e1})]:
\begin{equation}
\label{e6}
|\Psi\rangle = \frac{1}{\sqrt{2}} \left(  |1_{\omega_{-3}}\rangle_A  |1_{\omega_{4}}\rangle_B + |1_{\omega_{-4}}\rangle_A  |1_{\omega_{5}}\rangle_B \right).
\end{equation}
Ideally, parametric downconversion and filtering should produce this relative phase automatically; but in order to compensate any residual dispersion, we also fine-tune the phase with the BFC pulse shaper, experimentally maximizing spectral correlations in the $H_A \otimes H_B$ measurement case (see below). We have six initially empty modes between those populated in $A$ and $B$, allowing us to apply combinations of Hadamard operations and the identity to each pair of modes---$\{-4,-3\}$ and $\{4,5\}$---without any fear of the photon in $A$ jumping over to $B$'s modes, and vice versa (cf. guardband discussion in ref.~\onlinecite{Lu2018}). Accordingly, after the frequency-bin transformation $V$ (chosen to apply the desired joint operation), the coincidence probability for any ($n_A\leq0, n_B\geq 1$) is given by
\begin{equation}
\label{e7}
\nonumber
C_{n_A n_B} = \left|V_{n_A,-3} V_{n_B,4} + V_{n_A,-4} V_{n_B,5} \right|^2.
\end{equation}
This expression accounts for all aspects of the potentially nonideal mode transformation. Focusing on the qubit modes ($n_A\in\{-4,3\}, n_B\in\{4,5\}$), we have the ideal coincidences under all four cases of Fig.~\ref{fig4} as:
\begin{equation}
\label{e8}
\begin{aligned}
V(\mathbbm{1}_A \otimes \mathbbm{1}_B) & \Longrightarrow & C_{n_A n_B}^{\mathbbm{1}_A \otimes \mathbbm{1}_B} & = & \frac{1}{2} & \left( \delta_{n_A,-3}\delta_{n_B,4} + \delta_{n_A,-4}\delta_{n_B,5}  \right) \\
V(\mathbbm{1}_A \otimes H_B) & \Longrightarrow & C_{n_A n_B}^{\mathbbm{1}_A \otimes H_B} & = & \frac{1}{4} & \left( \delta_{n_A,-3}\delta_{n_B,4} + \delta_{n_A,-3}\delta_{n_B,5}+ \delta_{n_A,-4}\delta_{n_B,4} + \delta_{n_A,-4}\delta_{n_B,5}  \right) \\
V(H_A \otimes \mathbbm{1}_B) & \Longrightarrow & C_{n_A n_B}^{H_A \otimes \mathbbm{1}_B} & = & \frac{1}{4} & \left( \delta_{n_A,-3}\delta_{n_B,4} + \delta_{n_A,-3}\delta_{n_B,5} + \delta_{n_A,-4}\delta_{n_B,4} + \delta_{n_A,-4}\delta_{n_B,5}  \right) \\
V(H_A \otimes H_B) & \Longrightarrow & C_{n_A n_B}^{H_A \otimes H_B} & = & \frac{1}{2} & \left( \delta_{n_A,-4}\delta_{n_B,4} + \delta_{n_A,-3}\delta_{n_B,5}  \right),
\end{aligned}
\end{equation}
where $\delta_{nm} = \{1 \text{ if } n=m, 0 \text{ if } n\neq m\}$. These expressions predict perfect negative correlations for the case $\mathbbm{1}_A \otimes \mathbbm{1}_B$---i.e., detecting the low frequency of $A$ occurs in coincidence with the high frequency of $B$, and vice versa---while positive correlations result for $H_A \otimes H_B$. For the other two cases, no frequency correlations are present, with all four combinations equally likely.

The predictions of Eq.~(\ref{e8}) are precisely those of the EPR paradox~\cite{Einstein1935}, particularly the discrete version formulated by Bohm~\cite{Bohm1951}. Applying unitaries $\mathbbm{1}$ and $H$ followed by frequency detection produces measurements of the Pauli $Z$ and $X$ bases, respectively, so that our simultaneous correlations in two mutually unbiased bases reveals the strange characteristics of quantum mechanics, although admittedly without verifying, in our experiments, its nonlocal character. Finally, while we formulate the theory here viewing the operations as \emph{measurements}---as this is most direct in explaining our results---they can equivalently be considered as \emph{gates} on the input state, a useful distinction in considering more complex frequency networks where the photons are processed further before detection.

\textbf{Entanglement witness.} The EPR paradox exemplified by the theory of Eq.~(\ref{e8}) and results in Fig.~\ref{fig4} can be quantified in terms of an entropic entanglement witness between party $A$ and $B$. Adapting Eq.~(329) of ref.~\onlinecite{Coles2017} to our experiments gives following inequality, satisfied by all separable states:
\begin{equation}
\label{e9}
\nonumber
\mathcal{H}(\mathbbm{1}_A|\mathbbm{1}_B) + \mathcal{H}(H_A|H_B) \geq q_{MU},
\end{equation}
where $\mathcal{H}(U_A|U_B^\prime)$ is the entropy of $A$'s detected frequency bin, given knowledge of $B$'s result, when unitary operations $U$ and $U^\prime$ are applied. (We use the transformations as arguments, rather than the observables per se, for maximal clarity with our experimental configuration.) The Maassen-Uffink bound $q_{MU}$~\cite{Maassen1988} depends on the overlap between the frequency basis vectors in the rows of $\mathbbm{1_A}$ and $H_A$, which can be written in terms of the beamsplitter reflectivity $\mathcal{R}$ and transmissivity $\mathcal{T}$ (defined above) as
\begin{equation}
\label{e10}
\nonumber
q_{MU} = -\log_2 \max \left( \left\{ \frac{\mathcal{R}}{\mathcal{R+T}}, \frac{\mathcal{T}}{\mathcal{R+T}} \right\}\right),
\end{equation}
where $\{\mathcal{R},\mathcal{T}\}$ are evaluated at $\alpha=\pi$. Because we postselect on coincidences in the two-qubit Hilbert space, we have normalized $\{\mathcal{R},\mathcal{T}\}$ by the success probability; plugging in theoretical values, we obtain $q_{MU}=0.9710$---close to the maximum of 1 for perfect mutually unbiased bases in $d=2$ dimensions ($\log_2 d$).

To calculate the conditional entropies corresponding to the measurements in Fig.~\ref{fig4}, we employ Bayesian mean estimation (BME) on the raw count data~\cite{Blume2010, Williams2017}. Unlike alternative approaches, such as maximum likelihood estimation, BME yields error bars directly for any computed function and incorporates prior knowledge into the calculation naturally. To produce as conservative an estimate as possible, we make no specifying assumptions about the underlying state. For each situation in Fig.~\ref{fig4}, we posit a three-parameter multinomial likelihood function (four probabilities minus normalization), with counts taken directly from the raw data; we take the prior as uniform. The estimated means and standard deviations of the conditional entropies are then
\begin{equation}
\nonumber
\begin{aligned}
\mathcal{H}(\mathbbm{1}_A|\mathbbm{1}_B) & = & &0.19\pm 0.03 \\
\mathcal{H}(H_A|\mathbbm{1}_B) & = & &0.997\pm 0.003 \\
\mathcal{H}(\mathbbm{1}_A|H_B) & = & &0.993\pm 0.005 \\
\mathcal{H}(H_A|H_B) & = & &0.29\pm 0.04. \\
\end{aligned}
\end{equation}
As expected, the mismatched bases have near-maximal entropy (1 bit), while matched cases are much lower. We emphasize that the full effect of accidentals are included in these numbers; appreciably lower matched entropies may be possible in a model incorporating dark counts as well. Nonetheless, summing these entropies directly gives $\mathcal{H}(\mathbbm{1}_A|\mathbbm{1}_B) + \mathcal{H}(H_A|H_B) = 0.48 \pm 0.05$---violating the bound $q_{MU}$ by 9.8 standard deviations, and thereby confirming the nonseparability of our quantum state with high confidence.

\textbf{State reconstruction.} To estimate the complete two-qubit density matrix, we again employ BME~\cite{Blume2010, Williams2017} but now with the assumption of a single quantum state underlying all four measurements in Fig.~\ref{fig4}. As noted above, these four combinations are equivalent to joint measurements of the two-qubit observables $\{Z_A \otimes Z_B, X_A\otimes Z_B, Z_A\otimes X_B, X_A\otimes X_B \}$, where the identity $\mathbbm{1}$ permits measurement of $Z$, and the unitary $H$ allows measurement of $X$. As we have no information from the Pauli $Y$ basis, our set of projections is tomographically incomplete. We are nevertheless able to use the information from the measured bases to infer a complete state estimate, with appropriately higher uncertainties in the unmeasured bases. Finally, we emphasize that experimentally we only have access to the detector click (or no-click) events that are more naturally described in terms of positive-operator valued measures (POVMs) rather than von Neumann type projectors on the eigenvectors of Pauli $X$ and $Z$ operators.

For a specific two-qubit observable and chosen pair of frequency bins, we have the POVMs $\Lambda^{(A)}=\{\hat{\Pi}^{(A)}, \mathbbm{1}-\hat{\Pi}^{(A)}\}$ for subsystem $A$, and $\Lambda^{(B)}=\{\hat{\Pi}^{(B)}, \mathbbm{1}-\hat{\Pi}^{(B)}\}$ for subsystem $B$, where $\hat{\Pi}^{(A,B)}$ correspond to photon clicks, $\mathbbm{1}-\hat{\Pi}^{(A,B)}$ to the absence of a click. Absence of a click can be due to detection inefficiency or the photon being in an unmonitored mode. An outcome of a two-qubit POVM $\Lambda^{(A)} \otimes \Lambda^{(B)}$ will fall into one of the three experimentally recorded numbers: coincidence counts ($C_{AB}$), singles counts on detector $A$ ($S_A$), and singles counts on detector $B$ ($S_B$). These form our specific data set $\mathcal{D}=\{C_{AB}, S_A, S_B \}$. In our model, we assume fixed channel efficiencies for $A$ and $B$ propagation and detection ($\eta_A$ and $\eta_B$), and the following normalized probabilities under no loss and perfect detection: $p_{AB}$ (coincidence, one photon in mode $A$ and one photon in mode $B$), $p_{A0}$ (one photon in mode $A$ and no photon in mode $B$), $p_{0B}$ (one photon in mode $B$ and no photon in mode $A$), $p_{00}$ (no photon in mode $A$ or $B$).

 Letting $N$ denote the number of photon pairs generated in the measured time interval, we can enumerate the following four experimental possibilities, formed by the products of all operators from this POVM pair. (i) $\hat{\Pi}^{(A)}\otimes \hat{\Pi}^{(B)}$: coincidence between detectors $A$ and $B$. This occurs with probability $\eta_A \eta_B p_{AB}$ and is observed $C_{AB}$ times. (ii) $\hat{\Pi}^{(A)}\otimes [\mathbbm{1}-\hat{\Pi}^{(B)}]$: click on detector $A$, no click on $B$. This occurs with probability $\eta_A[p_{AB}(1-\eta_B) + p_{A0}]$ and is observed $S_A-C_{AB}$ times. (iii) $[\mathbbm{1}-\hat{\Pi}^{(A)}]\otimes \hat{\Pi}^{(B)}$: no click on detector $A$, click on detector $B$. This has probability $\eta_B [p_{AB}(1-\eta_A) + p_{0B}]$ and is observed $S_B-C_{AB}$ times. (iv) $[\mathbbm{1}-\hat{\Pi}^{(A)}]\otimes [\mathbbm{1}-\hat{\Pi}^{(B)}]$: no click on either detector. This occurs with probability $p_{AB}(1-\eta_A)(1-\eta_B) + p_{A0}(1-\eta_A) + p_{0B}(1-\eta_B) + p_{00}$ and is counted $N-S_A-S_B+C_{AB}$ times. Our likelihood function, $P(\mathcal{D}|\alpha)$, is then a multinomial distribution over the aforementioned probabilities and outcomes, where  $\alpha=\{\hat{\rho},\eta_A,\eta_B,N\}$ is the underlying parameter set. The idealized probabilities $\{p_{AB}, p_{A0}, p_{0B}, p_{00} \}$ are all functions of the density matrix $\hat{\rho}$, which we limit to physically allowable states~\cite{Williams2017}.

Up to this point, we have focused on a \emph{specific} choice of POVMs, $\Lambda^{(A)} \otimes \Lambda^{(B)}$. To account for all 16 POVM combinations (basis pairs and frequency-bin pairs) in the two-qubit space of Fig.~\ref{fig4}, we form the product over all settings, leaving the complete posterior distribution 
\begin{equation}
\label{b2}
P(\alpha|\bm{\mathcal{D}}) = \frac{\left[ \prod_j P(\mathcal{D}_j|\alpha)  \right] P(\alpha)}{P(\bm{\mathcal{D}})},
\end{equation}
where the bolded $\bm{\mathcal{D}}$ represents the union of the respective results $\mathcal{D}_j$ from each particular setting ($j=1,2,...,16$). Our prior $P(\alpha)$ is taken to be uniform in a Haar-invariant sense, and the marginal $P(\bm{\mathcal{D}})$ is found by integrating the numerator in Eq.~(\ref{b2}). With this posterior distribution, we can estimate any parameter of interest via integration, such as the mean density matrix
\begin{equation}
\label{b3}
\nonumber
\hat{\rho}_\mathrm{BME} = \int d\alpha \; P(\alpha|\bm{\mathcal{D}}) \hat{\rho}.
\end{equation}
Due to the complexity of integrals of this form, we employ numerical slice sampling for their evaluation~\cite{Neal2003,Williams2017}. The density matrix estimate is plotted in Extended Data Fig.~1. To highlight BME's natural treatment of error, we include the standard deviation for each matrix element as well. The error is extremely low for the real elements, due to our complete coverage of the $X$ and $Z$ bases, whereas several of the imaginary components (particularly the center two of the antidiagonal) have significantly higher error; this is a natural consequence of our lack of any measurements in the Pauli $Y$ basis. Nevertheless, physical requirements do bound this error, such that our Bayesian estimate of the fidelity, $\mathcal{F}=\langle\Psi|\hat{\rho}|\Psi\rangle$---with $|\Psi\rangle$ as defined in Eq.~(\ref{e6})---has extremely low uncertainty: $\mathcal{F}=0.92\pm0.01$. Such high fidelity provides positive corroboration of our frequency-bin control, and is fairly conservative, given that: (i) dark counts are not removed, and thus can degrade the state; and (ii) we lump any imperfections of our $H$ gate rotation onto the state itself, so that impurities in either the input state or quantum frequency processor will contribute to lower $\mathcal{F}$.

\newpage
\renewcommand{\thefigure}{1 (Extended Data)}
\begin{figure}
\centering
\includegraphics[width=4.5in]{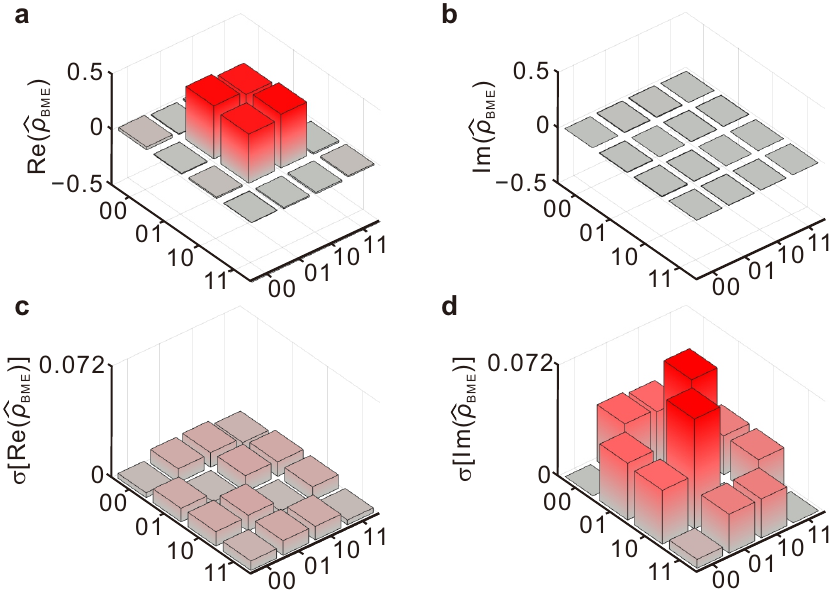}
\caption{\textbf{Reconstructed density matrix.} \textbf{a}, Real part of average density matrix. \textbf{b}, Imaginary part. \textbf{c}, Standard deviations of the real density matrix elements. \textbf{d}, Standard deviations of the imaginary elements. Label definitions: $00 \equiv |1_{\omega_{-4}}\rangle_A |1_{\omega_{4}}\rangle_B$, $01 \equiv |1_{\omega_{-4}}\rangle_A |1_{\omega_{5}}\rangle_B$, $10 \equiv |1_{\omega_{-3}}\rangle_A |1_{\omega_{4}}\rangle_B$, $11 \equiv |1_{\omega_{-3}}\rangle_A |1_{\omega_{5}}\rangle_B$.}
\label{figE1} 
\end{figure}

\newpage
\end{document}